\documentclass[twocolumn]{revtex4-2}
\usepackage[dvipdfmx]{graphicx}
\usepackage[usenames]{color}
\usepackage{amsmath,amsbsy,amssymb}
\usepackage{bm}
\usepackage{mathrsfs}
\usepackage{braket}
\usepackage{physics}
\usepackage{hyperref}
\usepackage{braket}
\usepackage{CJKutf8}

\graphicspath{{./figures/}}

\usepackage{color}

\newcommand{\diff}{\mathrm{d}}

\newcommand{\imu}{\mathrm{i}}
\newcommand{\epn}{\mathrm{e}}

\newcommand{\dg}{\dagger}
\newcommand{\la}{\langle}
\newcommand{\ra}{\rangle}
\newcommand{\al}{\alpha}
\newcommand{\sg}{\sigma}

\newcommand{\ep}{\varepsilon}

\newcommand{\rmII}{I\hspace*{-.1em}I}
\newcommand{\rmIII}{I\hspace*{-.1em}I\hspace*{-.1em}I}
\newcommand{\rmIV}{I\hspace*{-.1em}V}

\begin{document}

\title{\texorpdfstring{$\eta$}{[eta]}-pairing on bipartite and non-bipartite lattices}

\author{
Yutaro Misu$^{1}$, Shun Tamura$^{2}$, Yukio Tanaka$^{2}$ and Shintaro Hoshino$^{1}$
}

\affiliation{
$^1$Department of Physics, Saitama University, Saitama 338-8570, Japan \\
$^2$Department of Applied Physics, Nagoya University, Nagoya 464-8603, Japan
}

\date{\today}

\begin{abstract}
The \texorpdfstring{$\eta$}{[eta]}-pairing is a type of Cooper pairing state in which the phase of the superconducting order parameter is aligned in a staggered manner, in contrast to the usual BCS superconductors with a spatially uniform phase. In this study, we search for a characteristic \texorpdfstring{$\eta$}{[eta]}-pairing state in a triangular lattice where a simple staggered alignment of the phase is not possible. As an example, we consider the attractive Hubbard model on both the square and triangular lattices under strong external Zeeman field. Using the mean-field approximation, we have identified several \texorpdfstring{$\eta$}{[eta]}-pairing states. Additionally, we have examined the electromagnetic stability of the pairing state by calculating the Meissner kernel. Odd-frequency pairing plays a crucial role in achieving diamagnetic response if the electrons experience a staggered superconducting phase during the propagation of current.
\end{abstract}

\maketitle
\section{Introduction}
\label{sec:Intro}
The diversity of superconducting phenomena has been attracting continued attention. The superconducting state of matter is characterized by the properties of Cooper pairs, which can be classified based on their space-time and spin structures. With regard to their space structure, Cooper pairs are typically classified as $s$-wave, $p$-wave, or $d$-wave pairs depending on their relative coordinate structure. As for their center-of-mass coordinate, while it is usually assumed to be zero in most superconductors, it is possible to consider the existence of a finite center-of-mass momentum. One example of this is the Flude-Ferrell-Larkin-Ovchinnikov (FFLO) state \cite{Fulde1964,Larkin1964}, in which the Cooper pair has a small but finite center-of-mass momentum under the influence of a magnetic field. 
More generally, the magnitude of the center-of-mass momentum can be larger and of the order of the reciprocal lattice vector $\sim \pi/a$, where $a$ is a lattice constant. This type of pairing state is known as $\eta$-pairing, a concept first proposed by C. N. Yang, which forms a staggered alignment of the superconducting phase on a bipartite lattice \cite{Yang1989}. The spatially modulating order parameter is known also as the pair density wave, and has been discussed in relation to cuprate superconductors \cite{Agterberg20}.

The actual realization of the $\eta$-pairing has been proposed for the correlated electron systems such as the attractive Hubbard (AH) model with the magnetic field \cite{Rajiv1991}, the single- and two-channel Kondo lattices \cite{Coleman93, Hoshino_Kuramoto2014}, the Penson-Kolb model 
\cite{Wojciech2021}, and also the non-equilibrium situation \cite{Kaneko2019,Werner2019,Li2020,Nakagawa2021,Yang2022,Jiajun2022}. 
Since the phase of the superconducting order parameter can be regarded as the XY spin, the $\eta$-pairing is analogous to an antiferromagnetic state of the XY spin model.
Hence, the $\eta$-pairing state should be strongly dependent on the underlying lattice structure and we naively expect a variety of the $\eta$-pairing state if we consider the geometrically frustrated lattice such as the triangular lattice since the simple staggered state cannot be realized. 

In this paper, we deal with the AH model on the non-bipartite lattice in order to search for possible new superconducting states depending on the feature of the non-bipartite lattice structure in equilibrium. 
Already in the normal state without superconductivity, it has been pointed out that the non-bipartite lattice generates a non-trivial state of matter.
For example
in the Kondo lattice, a partial-Kondo-screening, which has a coexisting feature of Kondo spin-singlet and antiferromagnetism, is realized \cite{Motome2010}. 
Also in the AH model at half-filling, charge-density-wave (CDW) is suppressed due to the frustration effect \cite{Raimundo1993}. 
The $\eta$-pairing that appears in a photodoped Hubbard model on the triangular lattice has been studied recently \cite{Jiajun2022}. 
In the equilibrium situation, the properties of the AH model have been studied on bipartite lattices \cite{Rajiv1991}, but the model on a non-bipartite lattice has not been explored.

As shown in the rest of this paper, there are several types of $\eta$-pairings on the triangular lattice of the AH model under the Zeeman field.
 One of the $\eta$-pairing states is regarded as a 120$^\circ$-N\'eel state.
 Since the relative phase between the nearest neighbor sites is neither parallel nor anti-parallel, the inter-atomic Josephson current is spontaneously generated.
 This state can also be regarded as a staggered flux state, where the flux is created by the atomic-scale superconducting loop current.
While the staggered flux state has been studied so far \cite{Affleck1988, Chakravarty2001, Morr2002, Zhou2004, Yokoyama2016,Kenji2018, Fukuda2019}, the staggered flux in this paper is induced by the Josephson effect associated with superconductivity and has a different origin.

For the analysis of the AH model, we employ
the mean-field approximation in this paper.
It has been suggested that a simple $\eta$-pairing shows a paramagnetic Meissner state \cite{Hoshino2014}. 
Hence it is necessary to investigate the electromagnetic stability of the solution for superconductivity. 
We evaluate the Meissner kernel whose sign corresponds to the diamagnetic (minus) or paramagnetic (plus) response of the whole system, where the physically stable state should show diamagnetism.
We confirm that if the mean-field $\eta$-pairing state has the lowest energy compared to the other ordered states, the calculation of the Meissner kernel shows the diamagnetic response.
It is also notable that the odd-frequency pairing amplitude, which has an odd functional form with respect to the frequency \cite{Berezinskii74,Kirkpatrick91,Balatsky92,Coleman93,Tanaka12,Linder19,Cayao2020}, can contribute to the diamagnetism in the $\eta$-pairing state.
This is in contrast to the usual superconductivity with the uniform phase where the conventional even-frequency pairing contributes to the diamagnetism.
It has been shown that the odd-frequency pairing induced at the edge, interface or junctions \cite{Walter98,Tanaka05,Bergeret2005, Fomiunov2015,TanakaL2007,TanakaB2007} shows a paramagnetic response \cite{Higashitani97,Yokoyama11,Suzuki15,Bernardo15,Krieger20}.
In this paper, by contrast, we consider the odd-frequency pairing realized in bulk, which shows a qualitatively different behavior.

This paper is organized as follows.
We explain the model and method for the AH model in Sec.~\ref{sec:AH}, and the Meissner kernel in Sec.~\ref{sec:Meissner}.
The numerical results for the AH model are shown in Sec.~\ref{sec:Numerical result for AH}, and we summarize the paper in Sec.~\ref{sec:Summary}.
 
\section{
Attractive Hubbard model}
\label{sec:AH}
\subsection{Hamiltonian}
\label{sec:Model and Method for AH}

We consider the Hamiltonian of the AH model with magnetic field $\bm{h}$ which induce Zeeman effect only (Zeeman field) :
\begin{align}
{\cal{H}} = -t\displaystyle\sum\limits_{\langle i,j\rangle\sigma}c_{i\sigma}^{\dagger}c_{j\sigma}+\mathrm{H.c.} &+U\displaystyle\sum\limits_{i}n_{i\uparrow}n_{i\downarrow} \notag \\
& -\mu\displaystyle\sum\limits_{i}n_{i} -\bm{h}\cdot\displaystyle\sum\limits_{i}\bm{s}_{i}
\label{eq:Hamiltonian of AH},
\end{align}
where $c_{i\sigma}^{\dagger}$ and $c_{i\sigma}$ are the creation and annihilation operators of the $i$-th site with spin $\sigma$, respectively.
The symbol $\langle i, j\rangle$ represents a pair of the nearest-neighbor sites. Here, the parameter $t$ is the nearest-neighbor single-electron hopping integral. 
$U$ ($=-|U|$) is the onsite attractive interaction.
The spin operator is defined as $\bm{s}_{i}=\frac{1}{2}\sum_{\sigma\sigma^{\prime}}c_{i\sigma}^{\dagger}\bm{\tau}_{\sigma\sigma^{\prime}}c_{i\sigma^{\prime}}$, where $\bm{\tau}$ is the Pauli matrix, and the number operator of electrons is denoted as $n_{i}= n_{i\uparrow} + n_{i\downarrow} =\sum_{\sigma}c_{i\sigma}^{\dagger}c_{i\sigma}$.
The electron concentration is controlled by adjusting the chemical potential $\mu$.

\begin{figure}
    \centering
    \includegraphics[width=8.5cm]{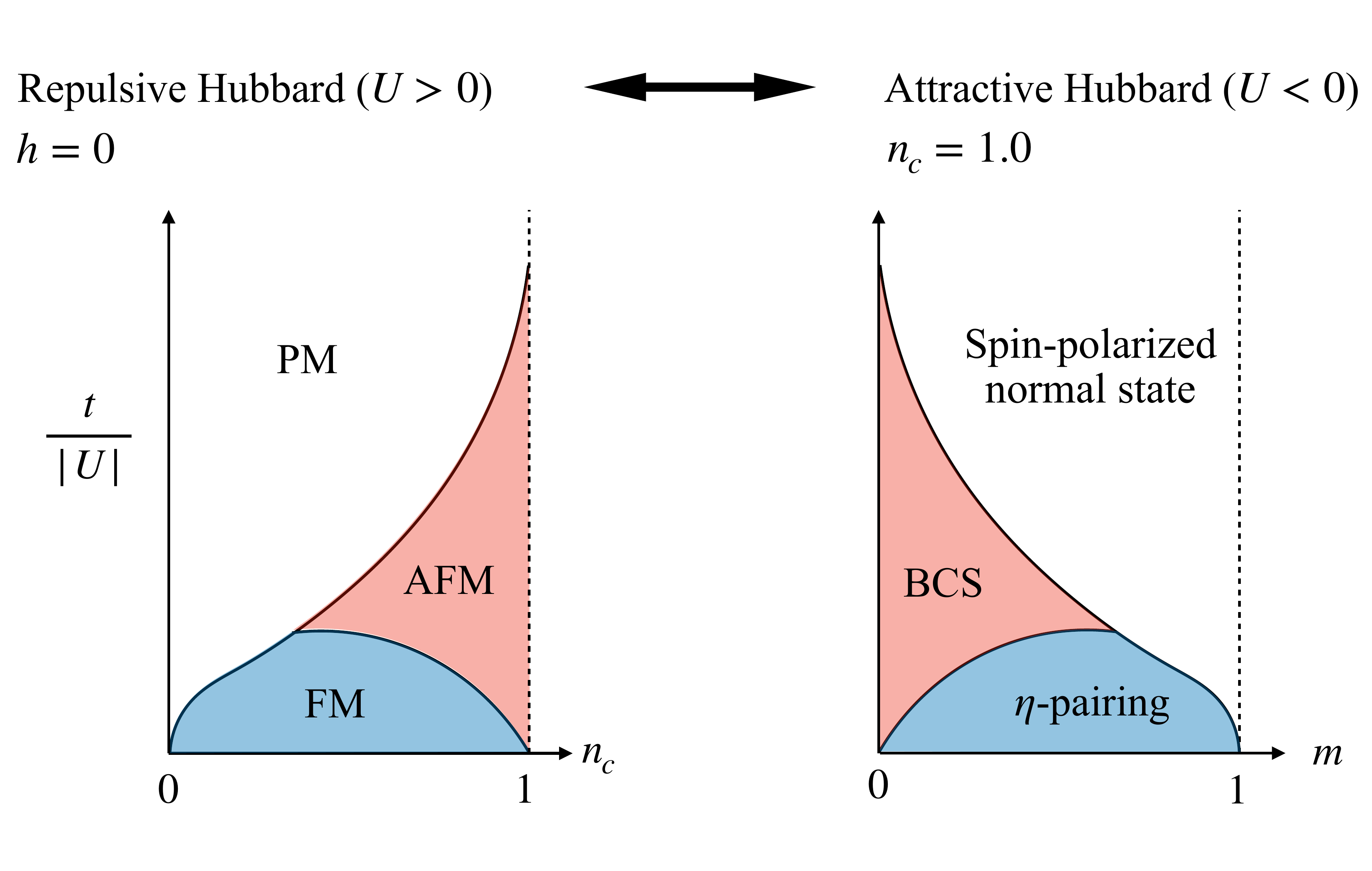}
    \caption{
    Sketches of the phase diagrams for the repulsive Hubbard model \cite{Claveau2014} (left panel) and AH model (right panel). $n_c$ is the electron concentration and $m$ is the magnetization. When the interaction $|U|$ is large, the ground state in the repulsive Hubbard model is ferromagnet (FM), while the ground state in the AH model is $\eta$-pairing. 
    }
    \label{fig:Repulsive Hubbard}
\end{figure}

The AH model has been successfully used to elucidate several important and fundamental issues in superconductors \cite{Micnas1990}.
The model on a bipartite lattice at half filling is theoretically mapped onto the repulsive Hubbard model by the following partial particle-hole transformation \cite{Shiba1972}
\begin{align}
    c_{i\uparrow}^{\dagger}\rightarrow ~c_{i\uparrow}^{\dagger},~c_{i\downarrow}^{\dagger} \rightarrow~c_{i\downarrow}\mathrm{e}^{{\rm{i}}\bm{Q}\cdot\bm{R}_{i}}.
    \label{eq:Shiba transformation}
\end{align}
The reciprocal vector $\bm{Q}$ satisfies the condition $\mathrm{e}^{{\rm{i}}\bm{Q}\cdot\bm{R}_{i}} = (-1)^{i}$ that takes $\pm 1$ depending on $\bm{R}_i$ belonging to A or B sublattice on the bipartite lattice.
Then, the $\eta$-pairing appears in the region that corresponds to a ferromagnet with transverse magnetization in the repulsive model \cite{Rajiv1991}. 
In a mean-field theory, the phase diagram for the repulsive Hubbard model without the magnetic field is shown in the left panel of Fig.~\ref{fig:Repulsive Hubbard} \cite{Claveau2014}. From this figure, we find that the ferromagnet is located in the regime where the repulsive interaction $U>0$ is large and the electron concentration is not half-filled. Hence, the $\eta$-pairing phase is located in the regime where the attractive interaction $U<0$ is large and the magnetization is finite. The phase diagram of the AH model at
half filling is shown in the right panel of Fig.~\ref{fig:Repulsive Hubbard}. 
In principle, an attractive interaction large enough to realize $\eta$-pairing could be realized in artificial cold atom systems \cite{Bloch2008}.

The Cooper pair is formed by the two electrons with~$(\bm{k} \uparrow,~-\bm{k} + \bm{q} \downarrow)$~where $\bm q$ is the center-of-mass momentum.
The FFLO state and the $\eta$-pairing are distinguished by the magnitude of $|\bm q|$.
In $\eta$-pairing, the center-of-mass momentum of the Cooper pair is the order of the reciprocal lattice vector, while the momentum of the FFLO state is much smaller and the spatial modulation is slowly-varying compared to the atomic scale.
Although the large center-of-mass momentum is usually not energetically favorable, a strong attractive interaction can make it  stable.

\subsection{Mean-field theory}

By applying the mean-field approximation, we obtain the mean-field  Hamiltonian
\begin{align}
{\cal{H}}^{\rm{MF}} &=-t\displaystyle\sum\limits_{\langle i,j\rangle\sigma}c_{i\sigma}^{\dagger}c_{j\sigma} +\mathrm{H.c.} -\mu\displaystyle\sum\limits_{i}n_{i} -\bm{h}\cdot\displaystyle\sum\limits_{i}\bm{s}_{i}\notag \\
&~~- \displaystyle\sum\limits_{i}\left(v_{i}n_{i} 
+\bm{H}_{i}\cdot\bm{s}_{i}- \Delta_{i}c_{i\uparrow}^{\dagger}c_{i\downarrow}^{\dagger} - \Delta^{*}_{i}c_{i\downarrow}c_{i\uparrow}\right).
\label{eq:MF_Hamiltonian_AH}
\end{align}
The order parameters are given by the self-consistent equations
\begin{align}
&v_{i} \equiv \dfrac{|U|}{2}\langle{n_{i}}\rangle
\label{eq:AH_CDW},\\
&\Delta_{i} \equiv -|U|\langle c_{i\downarrow}c_{i\uparrow}\rangle
\label{eq:AH_SC}, \\
&\bm{m}_{i} = \dfrac{1}{2}\displaystyle\sum\limits_{\sigma\sigma^{\prime}}\langle{c_{i\sigma}^{\dagger}\bm{\tau}_{\sigma\sigma^{\prime}}c_{i\sigma^{\prime}}}\rangle,~~\bm{H}_i= ~-2|U|\bm{m}_i,
\label{eq:AH_Spin}
\end{align}
where $\displaystyle \langle A \rangle = \mathrm{Tr}\left[A\mathrm{e}^{-{\cal{H}}^{{\rm{MF}}}/T}\right]
/ \mathrm{Tr}\left[\mathrm{e}^{-{\cal{H}}^{{\rm{MF}}}/T}\right]$ is a quantum statistical average with the mean-field Hamiltonian and $T$ is temperature.
$\Delta_i$ is the order parameter for $s$-wave singlet superconductivity (pair potential).
The phase $\theta_i \in [0,2\pi)$ of the pair potential $\Delta_i = |\Delta_i|\epn^{\imu \theta_i}$ is dependent on the site index and will be represented by the arrow in a two-dimensional space.
The mean-fields for the charge and spin are given by $v_i$ and $\bm H_i $, respectively, at each site.
The derivation of the self-consistent equations is summarized in Appendix~\ref{sec:appendix_self}.
We will consider the AH model both on the two-dimensional square and triangular lattices.

\section{Meissner kernel for a general tight-binding lattice}
\label{sec:Meissner}

\subsection{Definition}

As we explained in Sec.~\ref{sec:Intro}, it is necessary to calculate the Meissner kernel to determine whether the mean-field solution for $\eta$-pairing is electromagnetically stable. 
In the tight-binding model, the electromagnetic field appears as Peierls phase:
\begin{align}
    {\cal{H}}_{\rm{kin}}=-t\displaystyle\sum\limits_{\langle i,j\rangle\sigma}\mathrm{e}^{{\rm{i}}A_{ij}}c_{i\sigma}^{\dagger}c_{j\sigma} + \rm{H.c.}.
\end{align}
The Meissner effect is examined by the {\it weak} external orbital magnetic field applied perpendicular to the plane, while the $\eta$-pairing is stabilized only under a {\it strong} Zeeman field.
In order to make these compatible, 
we apply the Zeeman field parallel to the plane $\bm{h}=(h, 0, 0)$, which does not create the orbital motion of the tight-binding electrons.
Thus, the weak magnetic field that triggers the Meissner effect is applied perpendicular to the plane in addition to the in-plane magnetic field.
While the out-of-plane Zeeman effect is also induced by the weak additional field, it is neglected since the dominant Zeeman field already exists by the strong in-plane magnetic field.

Let us formulate the Meissner response kernel on a general tight-binding model.
We apply the formulation in Refs.~\cite{Scalapino92,Scalapino93,Kostyrko94} to the present case with sublattice degrees of freedom.
The current density operator between two sites is defined as 
\begin{align}
    \bm{j}_{ij}&=
    \dfrac{\partial {\cal{H}}_{\rm{kin}}}{\partial A_{ij}} \hat{\bm{\delta}}_{ij} \notag
    \\
    &=-\imu t\displaystyle\sum\limits_{\sigma}\left(c_{i\sigma}^{\dagger}c_{j\sigma}\mathrm{e}^{\imu A_{ij}}-c_{j\sigma}^{\dagger}c_{i\sigma}\mathrm{e}^{-\imu A_{ij}}\right)\hat{\bm{\delta}}_{ij},
\end{align}
where $\bm{\delta}_{ij} = \bm R_i - \bm R_j$ is the inter-site lattice vector between $i$-th and $j$-th sites, and hat ($\hat\ $) symbol means a unit vector.
In the linear response theory, the current operator which appears as a response to the static magnetic field in equilibrium is written as
\begin{align}
    \bm{j}_{ij}
    &\simeq -{\rm{i}}t\displaystyle\sum\limits_{\sigma} (c_{i\sigma}^{\dagger}c_{j\sigma}- c_{j\sigma}^{\dagger}c_{i\sigma})\hat{\bm{\delta}}_{ij} \notag \\
    &~~~~~~~+ t\displaystyle\sum\limits_{\sigma} (c_{i\sigma}^{\dagger}c_{j\sigma} +c_{j\sigma}^{\dagger}c_{i\sigma}) \hat{\bm{\delta}}_{ij}A_{ij} \notag \\
    &\equiv \bm{j}^{\rm{para}}_{ij} + \bm{j}^{\rm{dia}}_{ij}.
    \label{eq:current}
\end{align}
The first term is called the paramagnetic term and the second term is diamagnetic. 
The Fourier-transformed paramagnetic and diamagnetic current density operators are 
written as $\bm j^{\rm para}(\bm q)$ and $\bm j^{\rm dia}(\bm q)$.
{The linear response kernel is then defined by
$\langle{ j_{\nu}(\bm{q})}\rangle = \sum_{\mu}K_{\nu\mu}(\bm{q})A_{\mu}(\bm{q})$, where $\nu,\mu =x,y$ is the direction. 
We evaluate the kernel $K_{\nu\mu} (\bm{q}\rightarrow \bm{0}) \equiv K_{\nu\mu}$ when investigating the stability of superconductivity. 
This is called the Meissner kernel, which is proportional to the superfluid density. 

The Meissner kernel is separated into paramagnetic and diamagnetic terms as $K_{\nu\mu} = \left(K_{\rm{para}}\right)_{\nu\mu} + \left(K_{\rm{dia}}\right)_{\nu\mu}$. 
The paramagnetic kernel is given by
\begin{align}
\left(K_{{\rm{para}}}\right)_{\nu\mu} = \dfrac{1}{N}\displaystyle\int_{0}^{1/T}d\tau\langle j_{\nu}^{{\rm{para}}}(\bm{q}=0, \tau)j_{\mu}^{{\rm{para}}}(\bm{q}=0)\rangle,
\label{eq:K_para_df}
\end{align}
where $N = \sum_i 1$ is the number of sites.
The Heisenberg representation with the imaginary time $\tau$ is defined as $A(\tau) = \mathrm{e}^{{\cal{H}}\tau}A \mathrm{e}^{-{\cal{H}}\tau}$.
The form of the diamagnetic kernel is obvious from Eq.~\eqref{eq:current}.

We note that if the sign of the Meissner kernel $K$ is negative, the superconducting state is electromagnetically stable and is also called a diamagnetic Meissner state, which expels magnetic flux.
On the other hand, if the sign is positive, the superconducting state is called the paramagnetic Meissner state, which attracts magnetic flux.
For a stable thermodynamic superconducting state, the negative value of $K$ is required.

\subsection{Method of evaluation}

The actual evaluation of the kernels is performed based on the wave-vector representation.
Here, the physical quantities are described by the operator $c_{\bm k\sg}^{\al}$ where $\al$ distinguishes the sublattice.
Note that the Brillouin zone is folded by $\sum_\al 1$ times.
The diamagnetic kernel is rewritten as
\begin{align}
\left(K_{{\rm{dia}}}\right)_{\nu\mu} = \dfrac{1}{N}\displaystyle\sum\limits_{\alpha ,\beta}\displaystyle\sum\limits_{\bm{k}\sigma}\left(m^{-1}_{\bm{k}\alpha\beta}\right)_{\nu\mu}\langle{c_{\bm{k}\sigma}^{\alpha\dagger}c_{\bm{k}\sigma}^{\beta}}\rangle.
\label{eq: K_dia}
\end{align}
The inverse mass tensor $m^{-1}_{\bm{k}\alpha\beta}$, which reflects the characteristics of the lattice shape, are given by
\begin{align}
\left(m^{-1}_{\bm{k}\alpha\beta}\right)_{\nu\mu} \equiv t\displaystyle\sum\limits_{\langle i_\alpha ,j_\beta\rangle}\left({\hat{\bm{\delta}}}_{i_\alpha j_\beta}\right)_{\nu}\left({\hat{\bm{\delta}}}_{i_\alpha j_\beta}\right)_{\mu}\mathrm{e}^{-i\bm{k}\cdot\bm{R}_{{i}_\alpha {j}_\beta}},
\end{align}
where $i_\alpha$ is the $i$-th unit cell with} sublattice $\alpha$. The symbol $\langle i_\alpha ,j_\beta\rangle$ represents a pair of the nearest-neighbor sites and $\bm{R}_{{i}_\alpha {j}_\beta}$ is the vector between the unit lattice with the $i$-th sublattice $\alpha$ and the unit lattice with the $j$-th sublattice $\beta$.

The paramagnetic term has the form of a current-current correlation function.
We can calculate this term by using the Green's function matrix 
\begin{align}
    \check{\cal{G}}_{\bm{k}}(\tau) \equiv -\langle T_{\tau} \bm{\psi}_{\bm{k}}(\tau) \bm{\psi}_{\bm{k}}^{\dagger}\rangle
\end{align}
where $\bm{\psi}_{\bm{k}}=(c_{\bm{k}\uparrow}^{\alpha},c_{-\bm{k}\downarrow}^{\alpha\dagger}, \cdots)^T$ is the Nambu-spinor. 
$T_{\tau}$ is time-ordering operator regrading $\tau$. 
Each component of the Green's function matrix is given by the diagonal and off-diagonal Green's functions:
\begin{align}
    &G_{\sigma\sigma^{\prime}}^{\alpha\beta}(\bm{k},\tau) \equiv -\langle{T_{\tau}c_{\bm{k}\sigma}^{\alpha}(\tau)c_{\bm{k}\sigma^{\prime}}^{\beta\dagger}}\rangle, 
    \label{eq:G}\\
    &\bar{G}_{\sigma\sigma^{\prime}}^{\alpha\beta}(\bm{k},\tau) \equiv -\langle{T_{\tau}c_{\bm{k}\sigma}^{\alpha\dagger}(\tau)c_{\bm{k}^{\prime}\sigma^{\prime}}^{\beta}}\rangle,\\
    &F_{\sigma\sigma^{\prime}}^{\alpha\beta}(\bm{k},\tau) \equiv -\langle{T_{\tau}c_{\bm{k}\sigma}^{\alpha}(\tau)c_{-\bm{k}\sigma^{\prime}}^{\beta}}\rangle,
    \label{eq:pair amplitude}\\
    &F_{\sigma\sigma^{\prime}}^{\alpha\beta\dagger}(\bm{k},\tau) \equiv -\langle{T_{\tau}c_{-\bm{k}\sigma}^{\alpha\dagger}(\tau)c_{\bm{k}\sigma^{\prime}}^{\beta\dagger}}\rangle.
    \label{eq:F}
\end{align}
The anomalous part of Green's function [Eq.~(\ref{eq:pair amplitude})] is also called the pair amplitude.
The paramagnetic kernel in Eq.~\eqref{eq:K_para_df} can be divided into the normal ($G$) and anomalous ($F$) Green's function contributions as

\begin{widetext}

\begin{align}
\left(K_{\rm{para}}\right)_{\nu\mu} &= -\dfrac{1}{N}\displaystyle \sum\int_0^{1/T} \diff \tau \left(\bm{v}_{\bm{k}\alpha\beta}\right)_{\nu}\cdot\left(\bm{v}_{\bm{k}\alpha^{\prime}\beta^{\prime}}\right)_{\mu} 
\times\left(\bar{G}^{\alpha\beta^{\prime}}_{\sigma\sigma^{\prime}}(\bm{k},\tau)G^{\alpha^{\prime}\beta}_{\sigma\sigma^{\prime}}(\bm{k},\tau) + \bar{G}^{\alpha\beta^{\prime}}_{\sigma\sigma^{\prime}}(-\bm{k},\tau)G^{\alpha^{\prime}\beta}_{\sigma\sigma^{\prime}}(-\bm{k},\tau)\right)
\nonumber 
\\
&\hspace{3mm} -\dfrac{1}{N}\displaystyle \sum\int_0^{1/T} \diff \tau\left(\bm{v}_{\bm{k}\alpha\beta}\right)_{\nu}\cdot\left(\bm{v}_{-\bm{k}\alpha^{\prime}\beta^{\prime}}\right)_{\mu} 
\times \left(F^{\beta\alpha\dagger}_{\sigma^{\prime}\sigma}(\bm{k},-\tau)F^{\alpha^{\prime}\beta^{\prime}}_{\sigma,\sigma^{\prime}}(\bm{k},\tau) + F^{\beta\alpha\dagger}_{\sigma^{\prime}\sigma}(-\bm{k},-\tau)F^{\alpha^{\prime}\beta^{\prime}}_{\sigma,\sigma^{\prime}}(-\bm{k},\tau)\right) \notag 
\\
&\equiv K^G_{\rm{para}} + K^F_{\rm{para}}.
\label{eq:K_para}
\end{align}
The summation $\sum$ is performed over the indices which appears only in the right-hand side.
The velocity vector $\bm{v}_{\bm{k}\alpha\beta}$ is defined by
\begin{align}
    \left(\bm{v}_{\bm{k}\alpha\beta}\right)_{\nu}\equiv t\displaystyle\sum\limits_{\langle i_\alpha,j_\beta \rangle}\left(\hat{\bm{\delta}}_{i_\alpha j_\beta}\right)_{\nu}\mathrm{e}^{-
    \imu\bm{k}\cdot\bm{R}_{i_\alpha j_\beta}}.
\end{align}
In order to perform the integral with respect to $\tau$ in Eq.~(\ref{eq:K_para}), we define the Fourier-transformed Green's function as
\begin{align}
    g_{\bm{k}}({\rm{i}}\omega_n)\equiv \displaystyle\int_0^{1/T}d\tau g_{\bm{k}}(\tau)\mathrm{e}^{{\rm{i}}\omega_n\tau},
\end{align}
where $g_{\bm{k}}$ represents one of Eqs.~(\ref{eq:G})-(\ref{eq:F}) and $\omega_n=(2n + 1)\pi T$ is fermionic Mastubara frequency. Moreover, the Fourier-transformed Green's function matrix is given by using the matrix representation of mean-field Hamiltonian Eq.~(\ref{eq:MF_Hamiltonian_AH}) as 
\begin{align}
    \check{\cal{G}}_{\bm{k}}({\rm{i}}\omega_n)&=\left[{\rm{i}}\omega_n \check{1} - \check{\cal{H}}^{{\rm{MF}}}_{\bm{k}}\right]^{-1} =\check{U}_{\bm{k}}\left[{\rm{i}}\omega_n\check{1} - \check{\Lambda}_{\bm{k}}\right]^{-1}\check{U}^{\dagger}_{\bm{k}},
    \label{eq:Green's function with omega}
\end{align}
where $\check{\Lambda}_{\bm{k}}$ and $\check{U}_{\bm{k}}$ are, respectively, a diagonal eigenvalue matrix and a unitary matrix satisfying
$\check{U}^{\dagger}\check{\cal{H}}_{\bm{k}}^{{\rm{MF}}}\check{U} = \check{\Lambda}_{\bm{k}} = \mathrm{diag}(\lambda_{\bm{k}1},\lambda_{\bm{k}2},\ldots)$.
From Eq.~(\ref{eq:Green's function with omega}), $K_{\rm{para}}$ can be calculated as
\begin{align}
\left(K_{\rm{para}}\right)_{\nu\mu} &= -\dfrac{1}{N}\sum \left[\left(\bm{v}_{\bm{k}\alpha\beta}\right)_{\nu}\cdot\left(\bm{v}_{\bm{k}\alpha^{\prime}\beta^{\prime}}\right)_{\mu} 
 {\cal{U}}_{\bm{k}p}^{\beta^{\prime}\sigma^{\prime},\alpha\sigma}{\cal{U}}_{\bm{k}p^{\prime}}^{\alpha^{\prime}\sigma,\beta\sigma^{\prime}} 
 + \left(\bm{v}_{\bm{k}\alpha\beta}\right)_{\nu}\cdot\left(\bm{v}_{-\bm{k}\alpha^{\prime}\beta^{\prime}}\right)_{\mu} 
 {\cal{U}}_{\bm{k}p}^{\beta\sigma^{\prime},\alpha\sigma}{\cal{U}}_{\bm{k}p^{\prime}}^{\alpha^{\prime}\sigma,\beta^{\prime}\sigma^{\prime}}\right]
 \dfrac{f\left(\lambda_{\bm{k}p}\right) -f\left(\lambda_{\bm{k}p^{\prime}}\right)}{\lambda_{\bm{k}p}-\lambda_{\bm{k}p^{\prime}}} + \rm{c.c.}
 \label{eq:K_para_new}
\end{align}
where $f(\lambda_{\bm{k}p})=\frac{1}{\mathrm{e}^{ \lambda_{\bm{k}p}/T}+1}$ is the Fermi-Dirac distribution function and we have defined the coefficient ${\cal{U}}_{\bm{k}p}^{\alpha\sigma, \beta\sigma^{\prime}} \equiv \left[\check{U}_{\bm{k}}\right]_{\alpha\sigma,p}\left[\check{U}^{\dagger}_{\bm{k}}\right]_{p,\beta\sigma^{\prime}}$. 

\end{widetext}

The anomalous part of Eq.~(\ref{eq:K_para}) $K_{\rm{para}}^{F}$ is further decomposed into the contributions $K^{\rm{EFP}}$ and $K^{\rm{OFP}}$ from the even-frequency pair (EFP) and odd-frequency pair (OFP) amplitudes defined by 
\begin{align}
F^{\rm{EFP}}(\bm{k},\imu\omega_n) \equiv \dfrac{F(\bm{k},\imu\omega_n)+F(\bm{k},-\imu\omega_n)}{2}, 
\label{eq:F_even}
\\
F^{\rm{OFP}}(\bm{k},\imu\omega_n) \equiv \dfrac{F(\bm{k},\imu\omega_n)-F(\bm{k},-\imu\omega_n)}{2}.
\label{eq:F_odd}
\end{align}
Then, we obtain $K^{\rm{EFP}}$ and $K^{\rm{OFP}}$ by using Eqs.~(\ref{eq:F_even}) and (\ref{eq:F_odd}) as
\begin{align}
    K^{\rm{EFP,OFP}}_{\nu\mu} &= -\dfrac{1}{2N}\displaystyle\sum\limits_{\bm{k}}\displaystyle\sum\limits_{\alpha\beta\alpha^{\prime}\beta^{\prime}}\left(\bm{v}_{\bm{k}\alpha\beta}\right)_{\nu}\cdot\left(\bm{v}_{-\bm{k}\alpha^{\prime}\beta^{\prime}}\right)_{\mu} \notag \\
    &\times \displaystyle\sum\limits_{\sigma\sigma^{\prime}}\displaystyle\sum\limits_{pp^{\prime}}{\cal{U}}_{\bm{k}p}^{\beta\sigma^{\prime},\alpha\sigma}{\cal{U}}_{\bm{k}p^{\prime}}^{\alpha^{\prime}\sigma,\beta\sigma^{\prime}} \notag \\
    &\times \left[\dfrac{f\left(\lambda_{\bm{k}p}\right) -f\left(\lambda_{\bm{k}p^{\prime}}\right)}{\lambda_{\bm{k}p}-\lambda_{\bm{k}p^{\prime}}}
    \mp
    \dfrac{f\left(\lambda_{\bm{k}p}\right) -f\left(-\lambda_{\bm{k}p^{\prime}}\right)}{\lambda_{\bm{k}p}+\lambda_{\bm{k}p^{\prime}}} \right]  \notag \\
    & + \rm{c.c.},
\end{align}
where the minus ($-$) sign in the square bracket is taken for EFP contribution and the plus ($+$) for OFP pairing. These quantities are numerically calculated as shown in the next section.
Note that the cross term of the EFP and OFP terms of Green's functions vanishes after the summation with respect to the Matsubara frequency.

\subsection{Paramagnetic Meissner response of a simple \texorpdfstring{$\eta$}{[eta]}-pairing state}
Before we show the results of the AH model, let us show that a simple $\eta$-pairing state leads to the paramagnetic response which would not arise from thermodynamically stable states~\cite{Hoshino2014,Hoshino2016}.
We consider the simple bipartite lattice with staggered ordering vector $\bm Q$.
The anomalous contribution to the Meissner kernel may be written as \cite{Hoshino2016}
\begin{align}
    K_{{\rm para}, xx}^F &= -T\sum_{n\bm k\bm k'\sg\sg'}v_{\bm k}^x
 v_{\bm k'}^x 
 F^*_{\sg'\sg}(\bm k',\bm k,\imu\omega_n)
 F_{\sg\sg'}(\bm k,\bm k',\imu\omega_n).
\end{align}
This contribution must be negative (diamagnetic response) in order to dominate over the paramagnetic contribution.
For a purely $\eta$-pairing state, we assume the relation $F_{\sg\sg'}(\bm k,\bm k') = F_{\sg\sg'}(\bm k)\delta_{\bm k', -\bm k-\bm Q}$, and obtain
\begin{align}
    K_{{\rm para}, xx}^F &= -T\sum_{n\bm k\sg\sg'} (v_{\bm k}^x)^2
 F^*_{\sg'\sg}(\bm k,\imu\omega_n)
 F_{\sg\sg'}(\bm k,\imu\omega_n)
 ,
\end{align}
where we have used $v^x_{-\bm k-\bm Q} = v^x_{\bm k}$ valid for square lattice, which is in contrast to the relation $v^x_{-\bm k} = -v^x_{\bm k}$ for the uniform pairing with additional minus sign \cite{Hoshino2014}.
We separate the spin-singlet and triplet parts as $F_{\sg\sg'} = F_s \imu \tau^y_{\sg\sg'} + \bm F_t \cdot (\bm \tau \imu \tau^y)_{\sg\sg'}$, and then obtain
\begin{align}
    K_{{\rm para}, xx}^F &= 2T\sum_{n\bm k} (v_{\bm k}^x)^2
 \Big[ |F_{s}(\bm k,\imu\omega_n)|^2
 - |\bm F_{t}(\bm k,\imu\omega_n)|^2 \Big]
 .
\end{align}
If we consider the simple $\eta$-pairing with only spin-singlet part ($\bm F_t=\bm 0$), it leads to the paramagnetic response (positive).
\\
\indent 
Thus, a simple $s$-wave spin-singlet $\eta$-pairing is unlikely realized as a stable state.
On the other hand, in the AH model with magnetic field, the spin-triplet pair contribution is substantially generated by the Zeeman field, which plays an important role for the diamagnetic response as shown below.

\section{Numerical result for AH Model}
\label{sec:Numerical result for AH}

\subsection{Square lattice}
\label{sec:SQ in two-sublattice for AH} 

\subsubsection{Prerequisites}
Let us begin with the analysis of the AH model on the square lattice.
We consider the two-sublattice structure to describe the staggered ordered phase such as a $\eta$-pairing.
While the superconducting states in the attractive model are interpreted in terms of the magnetic phases of the repulsive model by the particle-hole transformation in Eq.~(\ref{eq:Shiba transformation}), the response functions such as the Meissner kernel are specific to the attractive model and have not been explored. 

In the following, we choose the band width $W=1$ as the unit of energy.
We fix the value of the attractive interaction $U=-1.375$.
The electron concentration is fixed as $n_c=1$, and the temperature is taken to be $T =1.0\times 10^{-3}$ unless otherwise specified.
We will investigate the change of the Meissner kernel for $\eta$-pairing as a function of magnetic field strength $h=|\bm{h}|$. 
In this paper, the mean-field solutions are calculated using the $60\times 60$ mesh in $\bm k$-space.
The result of the Meissner kernel for $\eta$-pairings is calculated with the mesh $300\times 300$. 
We also checked that the behaviors remain qualitatively unchanged when these numbers are increased.
The self-consistent equations in Eqs.~(\ref{eq:AH_CDW})-(\ref{eq:AH_Spin}) are computed by using an iterative method. In the following subsection \ref{sec:Two-sublattice solution}, we restrict ourselves to the analysis of two-sublattice mean-field solutions, and in \ref{sec:finite size square lattice for AH}, we examine the solutions when the two-sublattice constraint is relaxed.

\subsubsection{Two-sublattice solution}
\label{sec:Two-sublattice solution}
\begin{figure}
    \centering
    \includegraphics[width=8.5cm]{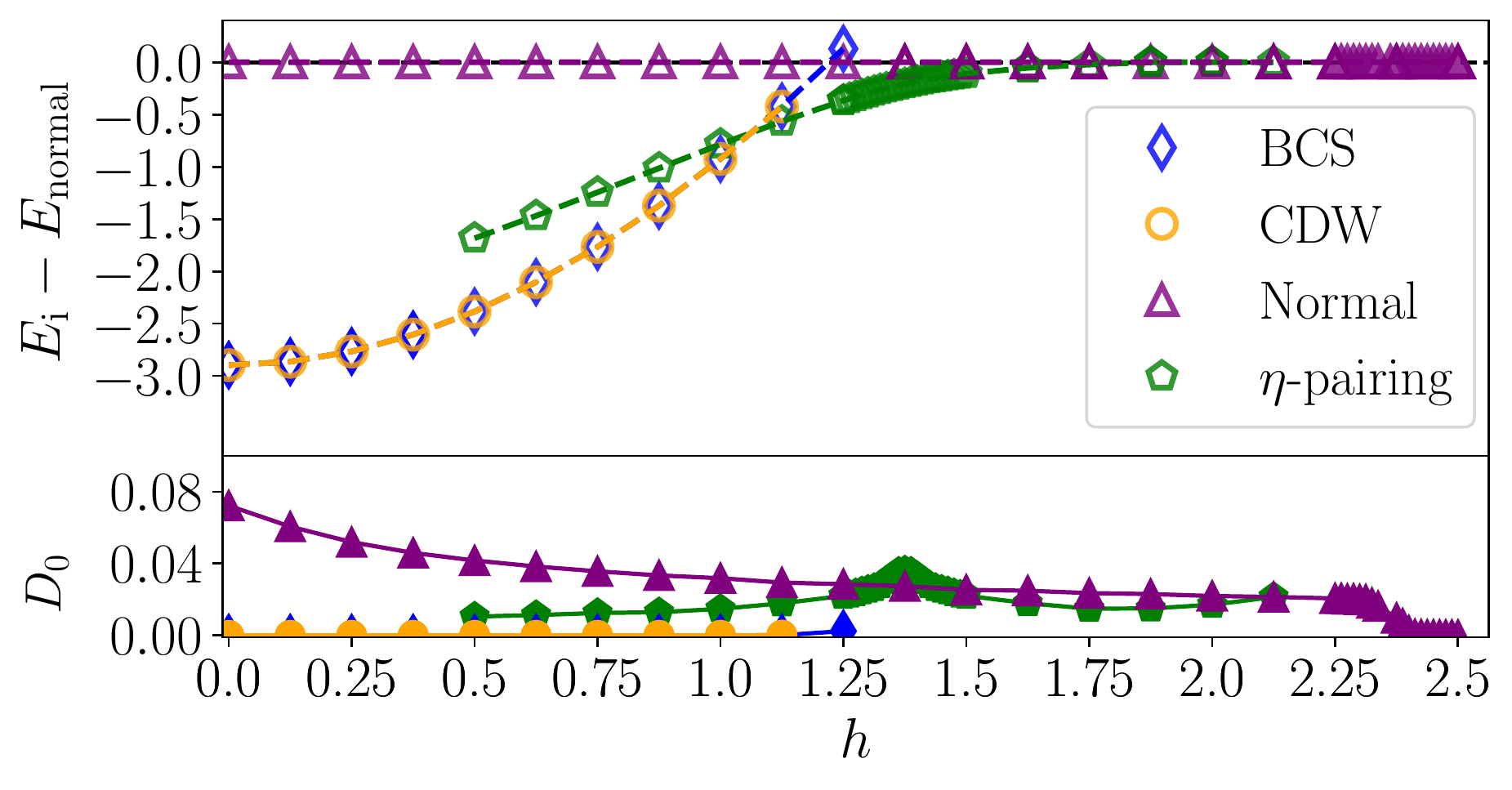}
    \caption{
    (Upper panel) Magnetic-field dependence of the internal energy for each state measured from the normal state in the square lattice model. 
    (Lower plane) Density of state (DOS) at zero energy $D_{0}$ for each state. 
    }
    \label{fig:SQ_phase}
\end{figure}

Before investigating the electromagnetic stability, we clarify the regime where the $\eta$-pairing becomes the ground state.
In this paper, we assume that the internal energy in Eq.~(\ref{eq:Hamiltonian of AH}) is approximately equal to the free energy in the low temperature region.
The upper panel of Fig.~\ref{fig:SQ_phase} shows the internal energy of several ordered states measured from the normal-state energy as a function of the Zeeman field $h$.
Here, the $\eta$-pairing solution is obtained by solving the self-consistent equation with imposing the constraint of the staggered phase of the pair amplitude.
A constraint is also used for the calculation of the other types of order parameters. 
Our calculations have not found any ordered states other than the types shown in Fig. \ref{fig:SQ_phase} even when a random initial condition is employed.

We determine the thermodynamically stable ground state by comparing the internal energies.
In low magnetic fields, BCS and CDW are degenerated ground states. 
On the other hand, we find that the $\eta$-pairing becomes the ground state in the magnetic field located in $1.063 < h <1.875$. 
The $\eta$-pairing solution itself is found in the wider regime although the internal energy is not the lowest one. 
It has been known that the attractive Hubbard model under a magnetic field also shows the FFLO state \cite{Tsuruta2014}, but this possibility cannot be considered when we take the two-sublattice condition.
This point will be revisited in the next subsection where the two-sublattice condition is relaxed.

\begin{figure}
\centering
\includegraphics[width=8.5cm]{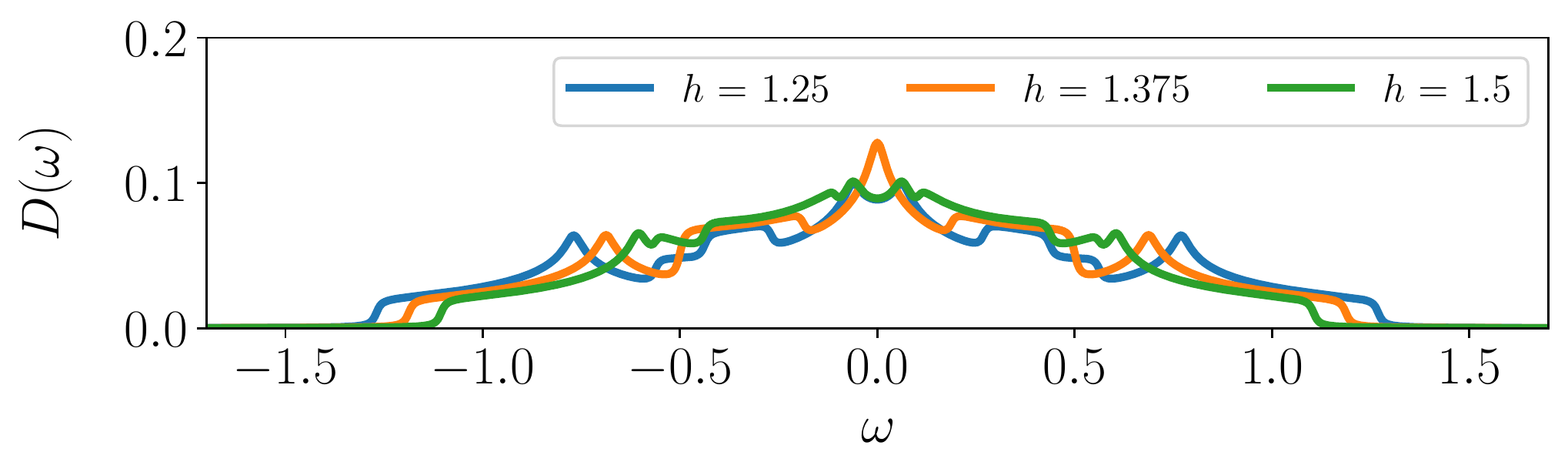}
\caption{Density of states for the $\eta$-pairing around magnetic filed $h=1.375$ in the square lattice model. Here $D(\omega)$ is normalized as $\int d\omega D(\omega) = 1$.
} 
\label{fig:DOS U =11}
\end{figure}

The lower panel of Fig.~\ref{fig:SQ_phase} shows the density of state~(DOS)~at the Fermi level for each state.
The result indicates that there is no energy gap in the $\eta$-pairing state,
in contrast to the conventional BCS pairing state.
There exists the regime where the DOS at the Fermi level for $\eta$-pairing is larger than that of normal metal ($1.25\lesssim h \lesssim 1.5$).
This is due to the van-Hove singularity of the square lattice model as shown in FIG.~\ref{fig:DOS U =11}.
We also perform the calculation for the cubic lattice where the van-Hove singularity is absent at zero energy and confirm in this case that the DOS is smaller than the normal state (see Appendix~\ref{sec:appendix_cubic}).

\begin{figure}
    \centering
    \includegraphics[width=8.5cm]{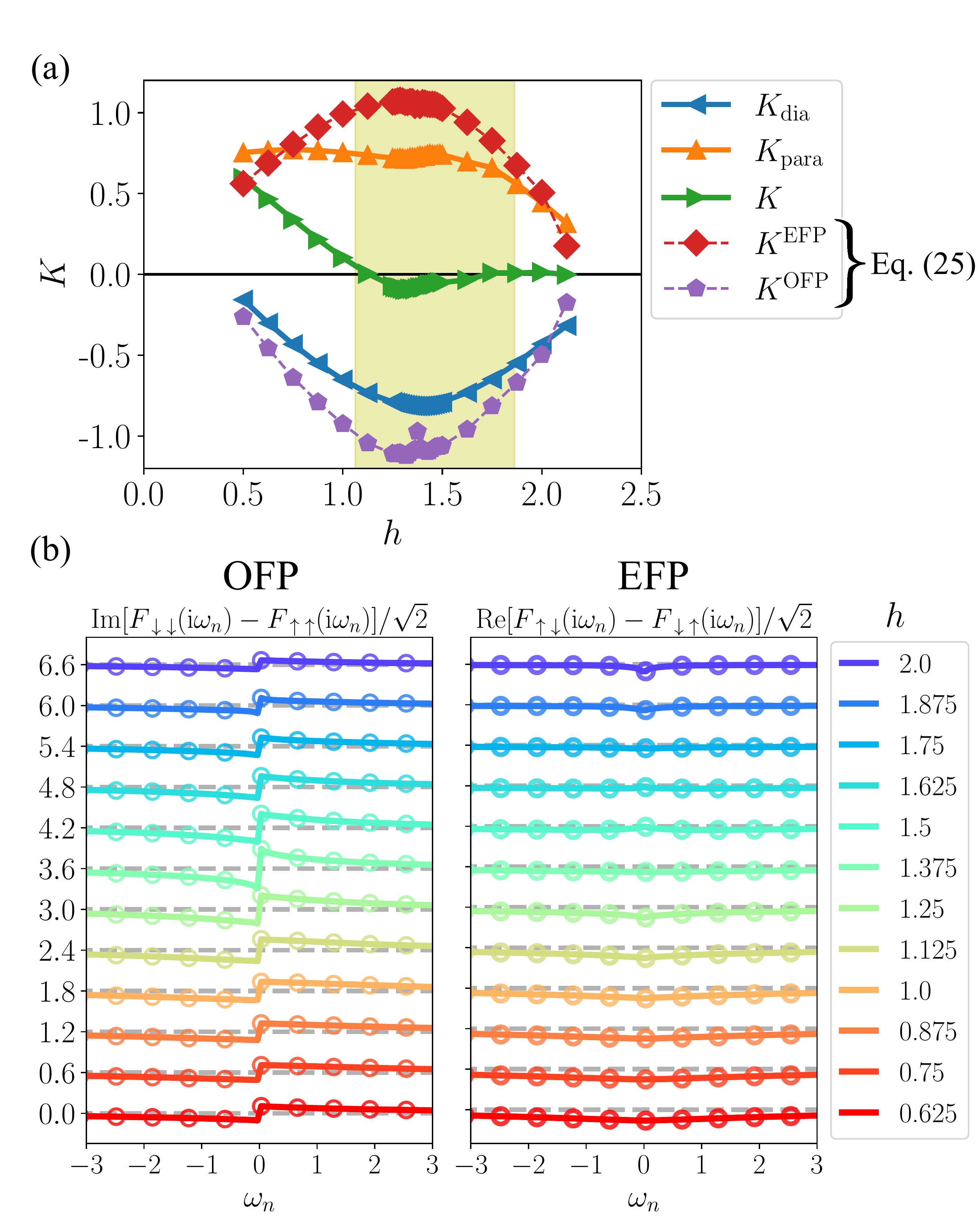}
    \caption{
    (a) Magnetic field dependence of the Meissner kernel $K(=K_{xx}=K_{yy})$ for the $\eta$-pairing on the square lattice. The yellow shaded rectangle indicates the range where the $\eta$-pairing becomes the ground state in two-sublattice calculation. 
    The number of the wavenumber $\bm{k}$ is taken as $300\times300$. 
    (b) Matsubara frequency dependence of the local pair amplitude at several magnetic fields. The left panel represents the imaginary part of $\left[F_{\downarrow\downarrow}(\imu \omega_n) - F_{\uparrow\uparrow}(\imu \omega_n)\right]/\sqrt{2}$, and the right panel represents the real part of $\left[F_{\uparrow\downarrow}(\imu \omega_n) - F_{\downarrow\uparrow}(\imu \omega_n)\right]/\sqrt{2}$. The values of the pair amplitudes are shifted by 0.6 at each magnetic field for visual clarity, and the gray-dotted lines are the zero axes for each magnetic field.}
    \label{fig:AH_Kernel_SQ}
\end{figure}

The stability of the $\eta$-pairing depends upon the magnitude of the magnetic field as seen in the Meissner response kernel $K$ ($=K_{xx}=K_{yy}$) (green symbol) in Fig.~\ref{fig:AH_Kernel_SQ}(a). 
The contributions from the paramagnetic ($K_{\rm{para}}$, positive) and diamagnetic ($K_{\rm{dia}}$, negative) parts are also separately plotted in the figure.
In the regime with $h \leq 1.125$ and $1.75 \leq h $, the $\eta$-pairing is electromagnetically unstable, while it is stable in $1.125 < h < 1.75$.
In Fig.~\ref{fig:AH_Kernel_SQ}, the yellow shaded rectangle indicates the regime where the $\eta$-pairing becomes the ground state as seen from Fig.~\ref{fig:SQ_phase}.
We find a narrow region where $\eta$-pairing is regarded as the ground state but is not an electromagnetically stable state around $h=1.125$.
From these results, we see that the $\eta$-pairing is not necessarily electromagnetically stable even if it becomes the ground state in a two-sublattice calculation.
As we shall see later, the simple $\eta$-pairing in this narrow regime does not necessarily exist if we relax the two-sublattice condition of the mean-field solution.

We also show in Fig.~\ref{fig:AH_Kernel_SQ}(a) the contributions from the even- and odd-frequency pairs defined in Eqs.~(\ref{eq:F_even}) and (\ref{eq:F_odd}).
The negative sign of the kernel, which means the response is diamagnetic, is partly due to the odd-frequency component of the pair amplitude, ($K^{\rm{OFP}}<0$).
This is in contrast to the FFLO state whose Meissner kernel is also negative due to the even-frequency component \cite{Debmalya2022}.
Hence, it implies that the mechanism of the diamagnetism is different between the FFLO and $\eta$-pairing states.

In addition to the Meissner kernel, we calculate the local pair amplitudes which are shown in FIG.~\ref{fig:AH_Kernel_SQ}(b). Here the left- and right-panels represent the spin-triplet and spin-singlet components of the local pair amplitude, respectively. 
The triplet component $\sum_{\sg\sg'} (\tau^\mu \imu\tau^y)_{\sg\sg'}F_{\sg\sg'}(\imu\omega_n)$ with $\mu=x$ has a finite imaginary part and zero real part, which represents the odd-frequency pair. 
The other $\mu = y,z$ components are zero.
On the other hand, the singlet component has a finite real part and zero imaginary part and is the even-frequency pair. 
We can see that the maximum value of the spin-triplet component of the pair amplitude is largest at the magnetic field $h=1.375$, where the magnitude of $K^{\rm{OFP}}$ is largest.
It is also notable that the magnitude of the odd-frequency pair amplitude correlates with the magnitude of DOS at zero energy as seen by comparing Figs.~\ref{fig:DOS U =11} and \ref{fig:AH_Kernel_SQ}.

We comment on the singular behavior of $K^{\rm{OFP}}$ at the magnetic field $h = 1.375$, although it does not affect the total Meissner kernel $K$. 
This anomalous feature is related to the van Hove singularity of the DOS at zero energy as shown in FIG.~\ref{fig:DOS U =11}, which shows a sharp peak at the Fermi level. 

\begin{figure}
    \centering
    \includegraphics[width=8.5cm]{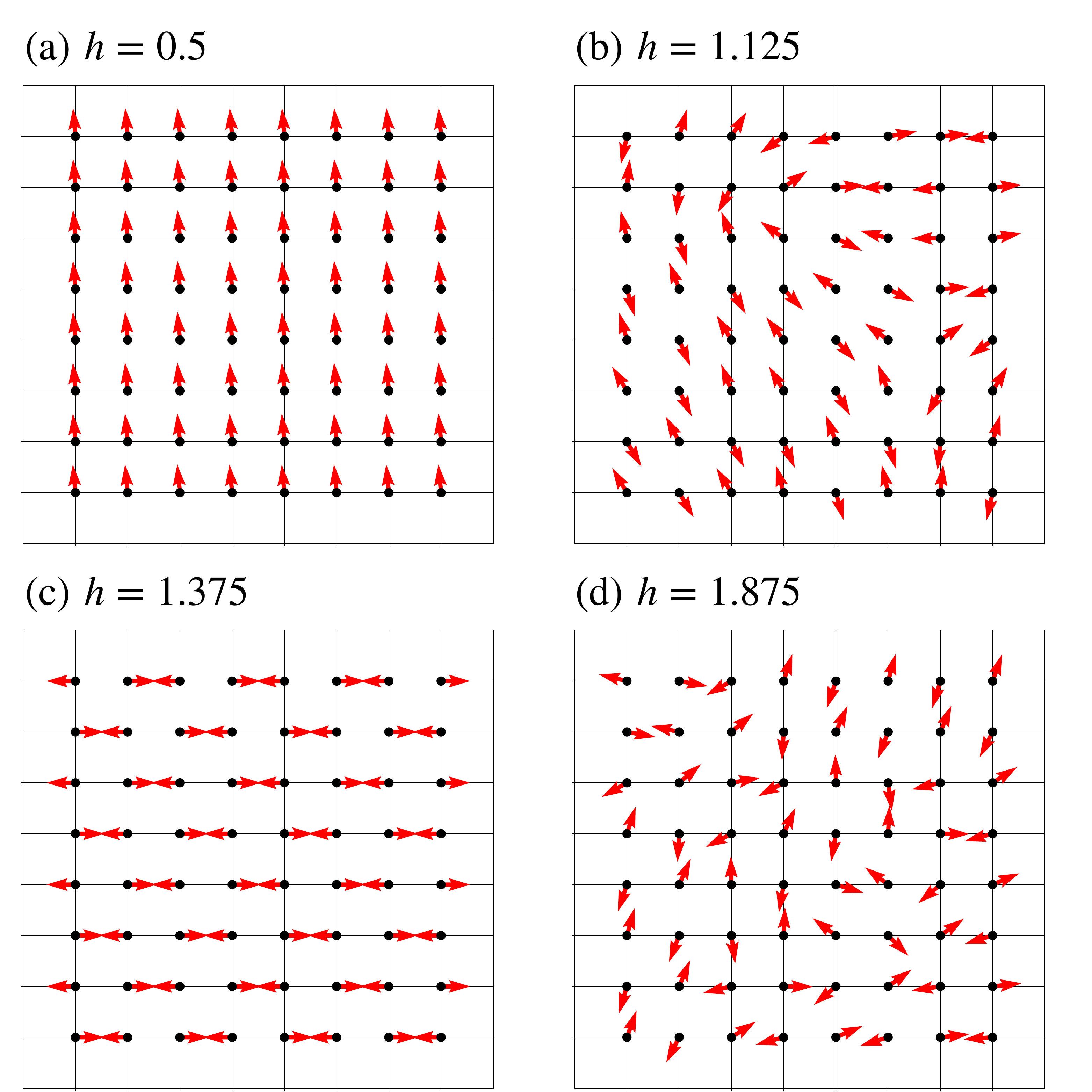}
    \caption{
    Spatial distribution of the phase of the superconducting order parameter at several magnetic fields.
    The calculation is performed on the finite-sized lattice $(8\times 8)$ with open boundary condition. Small black dots are lattice points and red arrows indicate the phase of the pair potential for each lattice point. 
    }
    \label{fig:U=11 Finite}
\end{figure}

\subsubsection{Beyond two-sublattice}
\label{sec:finite size square lattice for AH}

In order to clarify the stable ordered state where the Meissner kernel is positive (paramagnetic), we investigate mean-field solutions on finite-sized lattice where the two-sublattice condition is not imposed.
We have numerically solved the Eqs.~(\ref{eq:AH_CDW})-(\ref{eq:AH_Spin}) self-consistently by using the mean-field solutions of the $\eta$-pairing obtained for two-sublattice as an initial condition. 

Figure~\ref{fig:U=11 Finite} shows the spatial distribution of the phase of the gap function when the number of sites is $8\times 8$.
At $h=0.5$ in (a), where the $\eta$-pairing is not a ground state, the uniform BCS pairing state is realized as expected.
With increasing the magnetic field, the longer-periodicity structures are found as shown in Figs.~\ref{fig:U=11 Finite}(b), (c) and (d). 
At $h=1.375$ in (c), where the $\eta$-pairing solution has the lowest energy and the electromagnetic response is well diamagnetic, we obtain the staggered alignment of the phases.
When the parameters are close to the edges of the yellow-highlighted region in Fig.~\ref{fig:AH_Kernel_SQ}, the complex structures are formed as shown in (b) and (d).
The behavior in (b) is interpreted as due to the competing effect where the simple uniform and staggered phases are energetically close to each other. 

\begin{figure}
    \centering
    \includegraphics[width=8.5cm]{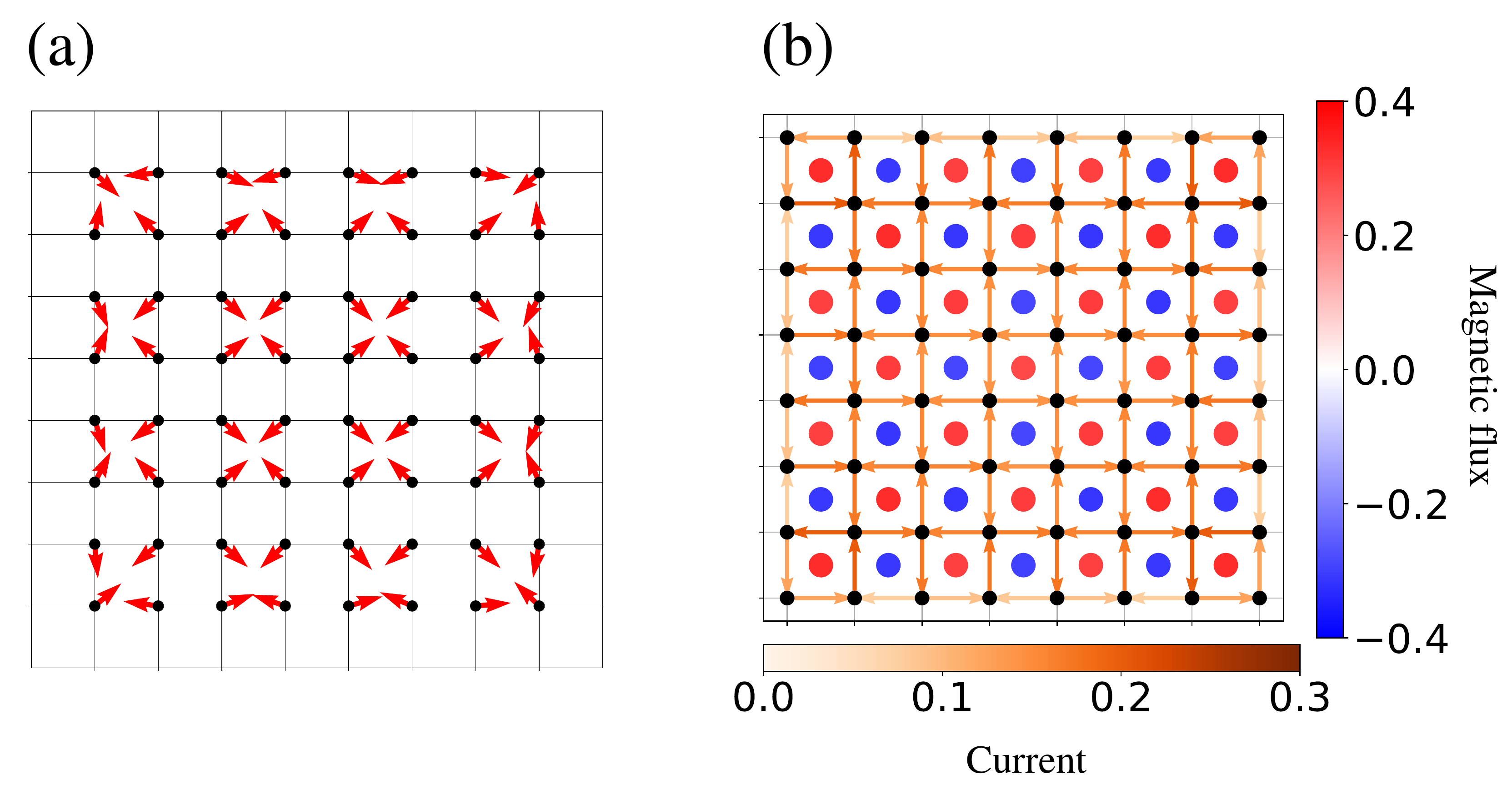}
    \caption{(a) 
    Spatial distribution of the phase of the superconducting order parameter for the $\eta$-pairing with 90$^{\circ}$-N\'eel state on the finite-sized lattice under open boundary conditions. (b) Spatial distributions of the spontaneous loop current and the flux defined on each plaquette. 
    The color of vectors displays the magnitude of current, and the color of dots in each plaquette indicates
    the value of the magnetic flux defined in Eq.~\eqref{eq:def_phi}.
    }
    \label{fig:loop_current_square_lattice}
\end{figure}

We also investigate the case with the other choice of parameters: $U = -1.25$ and $h =1.25$.
In this case, we find the staggered flux state where the phase of pair potential is characterized by 90$^\circ$-N\'eel ordering as in Fig.~\ref{fig:loop_current_square_lattice}(a). This ordered state cannot be described in the mean-field theory with two sublattices.
Owing to a non-colinear 90$^\circ$-N\'eel ordering vector, the spontaneous clockwise or counterclockwise loop currents arise by the inter-atomic Josephson effect.
The current density is calculated by
\begin{align}
    j_{ij} &= -{\rm{i}}t\displaystyle\sum\limits_{\sigma} \langle c_{i\sigma}^{\dagger}c_{j\sigma}- c_{j\sigma}^{\dagger}c_{i\sigma}\rangle
\end{align}
which is identical to the expression of the paramagnetic current in the linear response theory.
We can also evaluate the flux for each plaquette, which is define by
\begin{align}
    \Phi &= \sum_{(i,j)\in {\rm plaquette}} j_{ij}
    \label{eq:def_phi}
\end{align}
This expression is similar to the flux $\displaystyle \oint_C \bm j\cdot d\bm s = \int_S \bm b \cdot d \bm S$ ($\bm j = \bm \nabla \times \bm b$) defined in a continuum system, where $\bm{b}$ is a flux density. 
The flux is aligned in a staggered manner on a dual lattice as indicated in Fig. \ref{fig:loop_current_square_lattice}(b). 
The staggered flux originating from the normal part has been studied before \cite{Zhou2004, Yokoyama2016, Kenji2018,Fukuda2019}, while the staggered flux shown in Fig.~\ref{fig:loop_current_square_lattice}(b) has a different origin: it arises from the superconductivity associated with the off-diagonal part in the Nambu representation.

We also comment on a feedback effect to the electromagnetic field from the supercurrent. Since the characteristic length scale for the magnetic field in layered superconductor becomes long \cite{Pearl64}, each magnetic flux on the plaquette is smeared out with this length.
Hence we expect that the net magnetic field is not created from the staggered superconducting flux.

\subsection{Triangular lattice}
\label{sec:triangular lattice for AH}

\subsubsection{Mean-field solution}

Now we search for the $\eta$-pairing reflecting the characteristics of a geometrically frustrated triangular lattice at the half-filling ($n_c = 1.0$).
We choose the parameters $U=-1.83$ and $T=1.0\times10^{-3}$.
We consider the cases of two- and three-sublattice structures.
For a usual antiferromagnet, the typical ordered state in the two-sublattice case has a stripe pattern, while in the three-sublattice case we expect a 120$^\circ$-N\'eel state.
Below we study the superconducting $\eta$-pairing phases within the mean-field theory.

\begin{figure}[t]
    \centering
    \includegraphics[width=8.5cm]{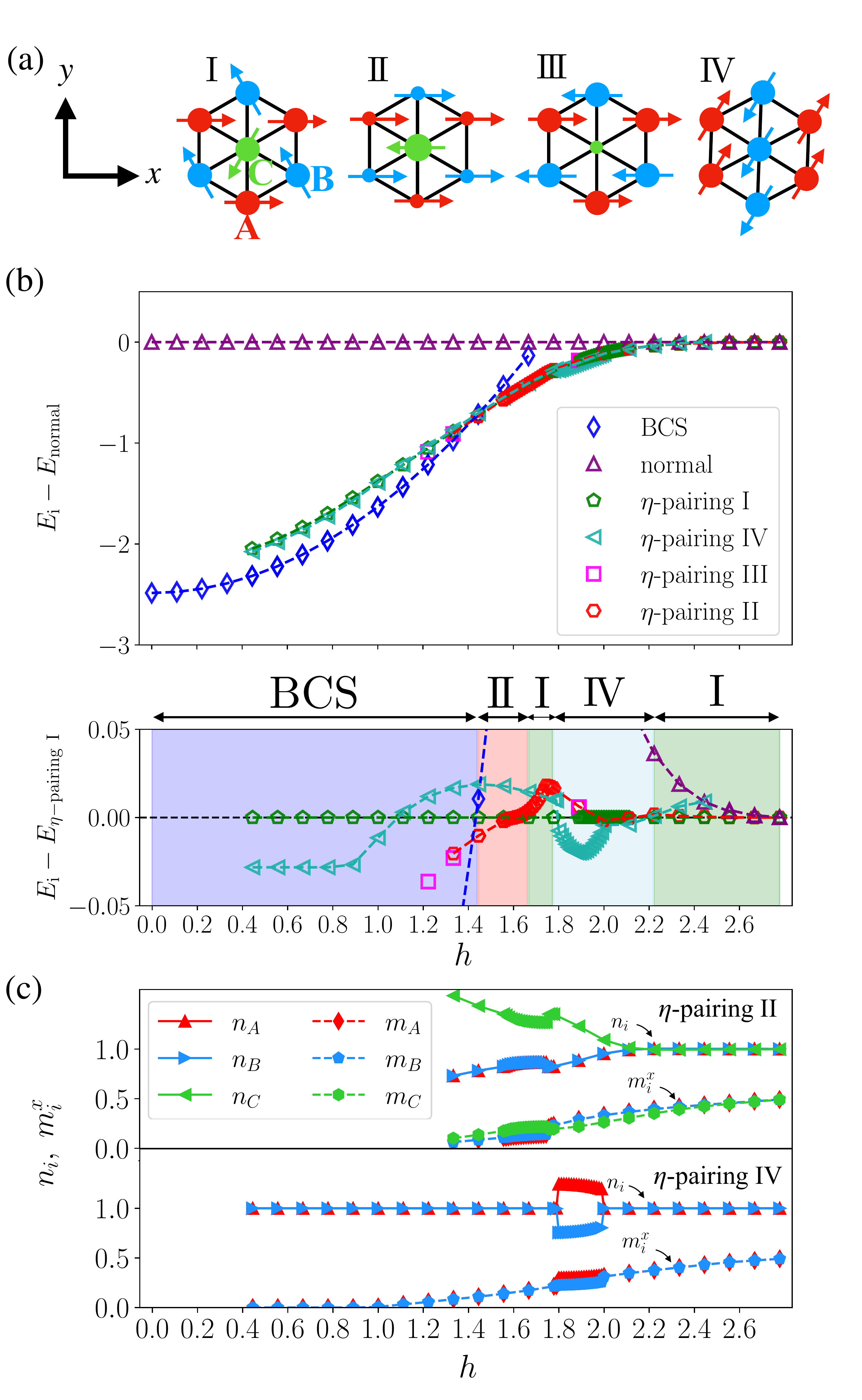}
    \caption{
    (a) Schematics for the four $\eta$-pairings in the triangular lattice model. The arrows indicate the phase of the pair potential. The size of the circles shows the amount of the electron density for each sublattice. 
    (b) Magnetic field dependence of the internal energies measured from the normal state (upper panel). The lower panel shows the internal energy measured from the $\eta$-pairing I. (c) Magnetic field dependence of the number of electrons and magnetization on each sublattice for the $\eta$-pairing {\rmII} (upper panel) and {\rmIV} (lower panel).}
    \label{fig:AH phase diagram No1}
\end{figure}

We have found the four types of superconducting states reflecting the characteristics of the triangular lattice, which are referred to as the $\eta$-pairing I, \rmII, \rmIII, and \rmIV. The schematic pictures for these four states are shown in Fig.~\ref{fig:AH phase diagram No1}(a),
where the arrow indicates the phase of the superconducting order parameter at each site.
We make a few general remarks:
the three-sublattice structure is assumed for \rm{I}, \rmII, \rmIII, while the two sublattice is employed for \rmIV.
The type-I has a non-colinear structure, and in the other $\eta$-pairings the vectors are aligned in a colinear manner.
We also note that CDW accompanies the $\eta$-pairings {\rmII} and {\rmIII}, where the number of local filling is indicated by the size of the filled circle symbols in Fig.~\ref{fig:AH phase diagram No1}(a).

Figure~\ref{fig:AH phase diagram No1}(b) shows the internal energy of the ordered states measured from the normal state (Upper panel) and from the $\eta$-pairing I (Lower panel). From the lower panel of Fig. \ref{fig:AH phase diagram No1}(b), we can identify the ground state. 
With increasing the magnetic field, the ground state changes as BCS $\rightarrow$ $\eta$-pairing \rmII $\rightarrow$ $\eta$-pairing I $\rightarrow$ $\eta$-pairing \rmIV $\rightarrow$ $\eta$-pairing I $\rightarrow$ normal.
Figure \ref{fig:AH phase diagram No1}(c) shows the particle density and $x$-direction magnetization $m_i^x$ of each sublattice for $\eta$-pairing {\rmII} (Upper panel) and $\eta$-pairing {\rmIV} (Lower panel). The values of $m_i^y$and $m_i^z$ are zero because the Zeeman field $\bm{h}$ is applied along the $x$-direction. Below, we explain the characteristic features for each $\eta$-pairing state.

\begin{figure}[t]
    \centering
    \includegraphics[width =8.5cm]{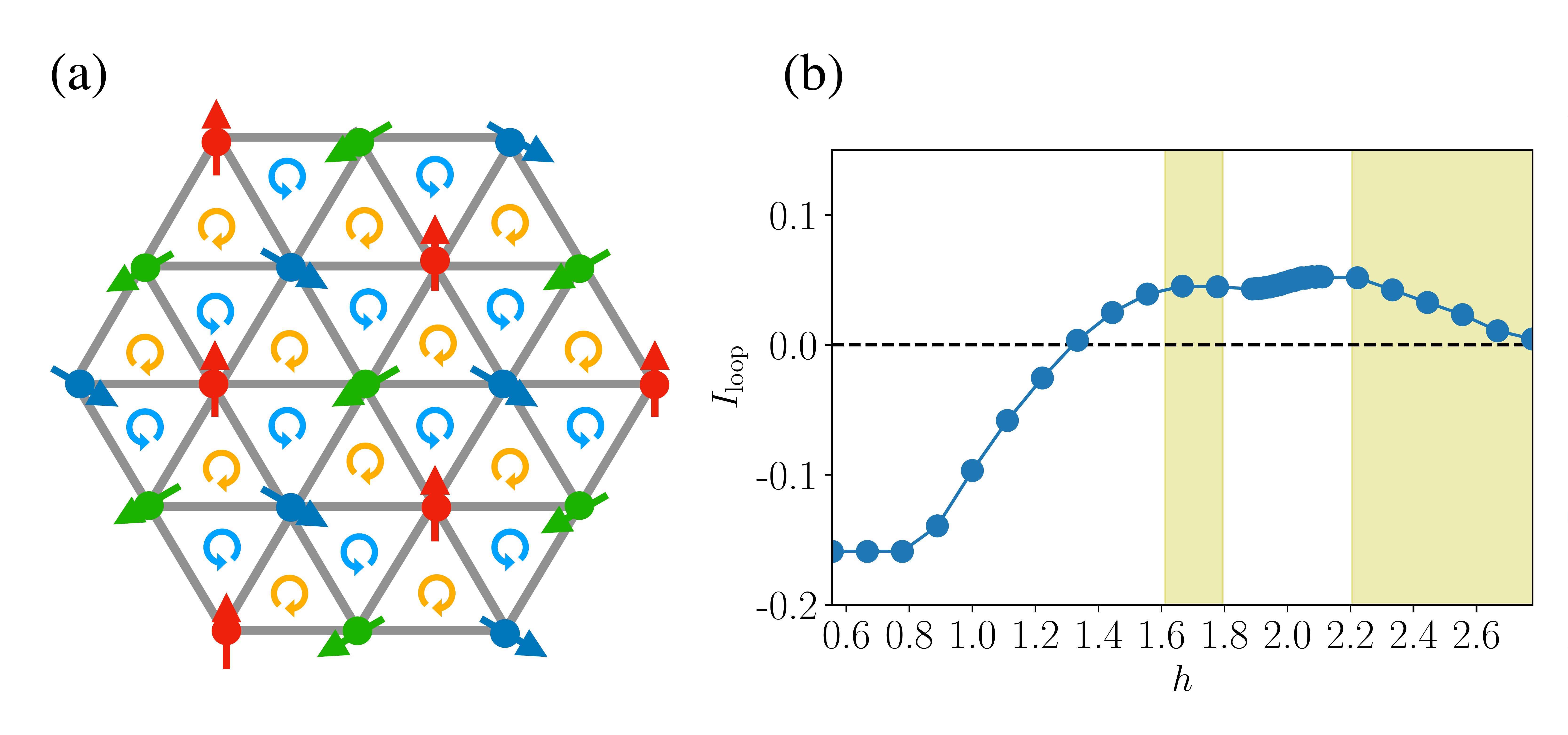}
    \caption{(a) Schematic picture of the staggered flux state on the triangular lattice. The straight arrows display the phase of the pair potential at each site, and the circle arrows indicate the staggered loop current. (b) Magnetic field dependence of the magnitude of loop current. The yellow shaded rectangle indicates the range where the $\eta$-pairing I becomes the ground state. }
    \label{fig:loop_current}
\end{figure}    

{\it $\eta$-pairing-I state.---}
The $\eta$-pairing I has 120$^\circ$ N\'eel ordering vector (Green pentagon in Fig. \ref{fig:AH phase diagram No1}(b)).
The spontaneous supercurrent appears in this non-colinear state as schematically shown in Fig.~\ref{fig:loop_current}(a).
This superconducting state forms a staggered flux state, where the flux is aligned on a honeycomb dual lattice, which
is similar to the $\eta$-pairing with 90$^\circ$-N\'eel ordering vector on the square lattice shown in Fig.~\ref{fig:loop_current_square_lattice}(b). 
Figure \ref{fig:loop_current}(b) displays the values of spontaneous loop current density as a function of the magnetic field. 

{\it $\eta$-pairing-{\rmII} state.---}
The $\eta$-pairing {\rmII} has the structure with up-up-down colinear phases
plus CDW (Red hexagon in Fig.~\ref{fig:AH phase diagram No1}(b)).
There is the relation $n_{\rm{A}} = n_{\rm{B}} <n_{\rm{C}}$ for the electron filling at each sublattice shown in Fig.~\ref{fig:AH phase diagram No1}(c).
We note that this site-dependent feature is characteristic for the {\rmII} (and \rmIV) state.
The phases of the pair potential at A and B sublattices 
are ``ferromagnetic'', while the phase at C sublattice  is ``antiferromagnetic''. The resulting ordered state is regarded as the emergence of the honeycomb lattice formed by equivalent A and B sublattices.

{\it $\eta$-pairing-{\rmIII} state.---}
This is the $\eta$-pairing with a staggered ordering vector and CDW (Magenta square in Fig.~\ref{fig:AH phase diagram No1}(b)).
The order parameter $\Delta$ at C sublattice is zero, but the others (A,B) are finite. 
The electron-rich sublattices A and B form a simple bipartite $\eta$-pairing state on an emergent honeycomb lattice.
Since this state does not become a ground state anywhere for the present choice of $U=-1.83$, we do not further investigate this state in the following.

{\it $\eta$-pairing-{\rmIV} state.---}
This is the $\eta$-pairing with  a simple stripe alignment (Cyan rhombus in Fig.~\ref{fig:AH phase diagram No1}(b)).
This $\eta$-pairing is accompanied by CDW around $h = 1.9$ shown in Fig.~\ref{fig:AH phase diagram No1}(c). 
As shown below, this stripe phase show an anisotropic behavior in linear response coefficients, while the other $\eta$-pairing states are isotropic.

\subsubsection{Meissner response}

\begin{figure}[t]
    \centering
    \includegraphics[width=8.5cm]{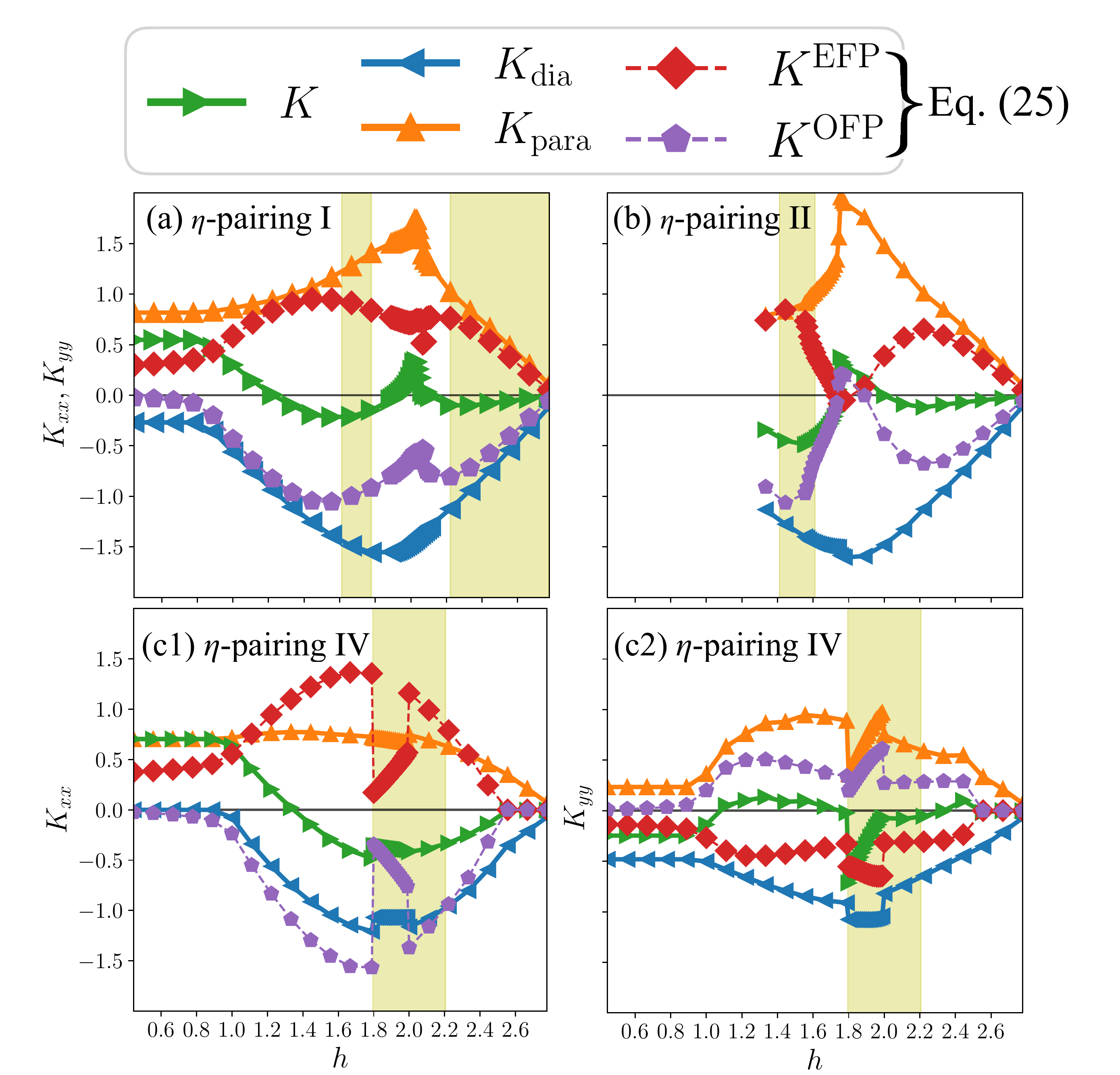}
    \caption{Magnetic field dependence of the Meissner kernels $K_{xx}$ and $K_{yy}$ for the $\eta$-pairings I, {\rmII}, {\rmIV} on the triangular lattice.
    The yellow shaded rectangle indicates the regime where each $\eta$-pairing becomes the ground state. The symbols are the same as those in Fig. \ref{fig:AH_Kernel_SQ}(a).
    For the $\eta$-pairing IV, $K_{xx}$ and $K_{yy}$ are separately plotted in (c1) and (c2).
    }
    \label{fig:TR_Kernel}
\end{figure}

Now we discuss the Meissner response.
Figure~\ref{fig:TR_Kernel}(a,b,c) shows the Meissner kernels $K_{xx},~K_{yy}$ for the $\eta$-pairing I, {\rmII} and {\rmIV}. 
The yellow-highlighted parts indicate the region where each $\eta$-pairing becomes the ground state as identified from Fig.~\ref{fig:AH phase diagram No1}(b).
The result for the $\eta$-pairing {\rmIII} is not shown because it does not become a ground state at $U=-1.83$.
We confirm that the Meissner response is basically diamagnetic if the $\eta$-pairing becomes the ground state as shown in Figs.~\ref{fig:TR_Kernel}(a,b,c). Thus the energetic stability and diamagnetic response are reasonably correlated. 
In the following, we discuss the properties of the Meissner kernel for each state.

The Meissner kernels for both $\eta$-pairing I and $\eta$-pairing {\rmII} shown in Figs.~\ref{fig:TR_Kernel}(a) and (b) satisfy the relation $K_{xx}=K_{yy}$, which means an isotropic linear response.
For the $\eta$-pairing I, the Meissner kernel becomes positive in the regions $h<1.2,~1.95<h<2.12$, while the kernel becomes negative in the ground state region (Fig. \ref{fig:TR_Kernel}(a)). Although the local current density is finite for the $\eta$-pairing I state, it does not affect the expression of the Meissner kernel in Eq.~\eqref{eq:K_para_df} since the total current $\bm j(\bm q=0)$ is zero. 

Next we disucuss the $\eta$-pairing IV state.
The Meissner kernel jumps at $h =1.8$ due to the emergence of the CDW order parameter as shown in Fig.~\ref{fig:TR_Kernel}(c1,c2). 
It is notable that the $\eta$-pairing IV with the stripe pattern shows a difference between $x$ and $y$ directions as shown in Figs.~\ref{fig:TR_Kernel}(c1,c2), respectively.
This characteristic behavior can be intuitively understood from Fig.~\ref{fig:AH phase diagram No1}(a), where the current along the $x$-axis flows with experiencing a staggered pair potential, whereas the current in the $y$-direction feels an uniform pair potential. 
In the Meissner response, $K_{xx}$ shows a characteristic behavior of the $\eta$-pairing, while $K_{yy}$ is qualitatively the same as the kernel of BCS. 
Thus, as shown in Fig.~\ref{fig:TR_Kernel}(c1), the diamagnetic response in the $x$-axis direction is related to to the odd-frequency pair, whereas the diamagnetic response in the $y$-axis direction, shown in Fig. \ref{fig:TR_Kernel}(c2), is related to even-frequency pair. 

\section{Summary and Outlook}
\label{sec:Summary}

We have studied the square and the triangular lattice of the attractive Hubbard model by using the mean-field theory.
Several types of $\eta$-pairing have been found in the triangular lattice where a simple bipartite pattern is not allowed.
Using the formulation of the Meissner kernel for a general tight-binding lattice, we have investigated the electromagnetic stability of $\eta$-pairings. 
We have confirmed that the electromagnetic stability of the $\eta$-pairing correlates with the internal energy.
In a narrow parameter range, we also find that the $\eta$-pairing state can show an unphysical paramagnetic response if we assume the two or three sublattice structure in the mean-field calculation.
In this case, another solution with longer periodicity needs to be sought.

When the current path experiences the staggered phase of the superconducting order parameter, the odd-frequency component of the pair amplitude contributes to the diamagnetic response.
This is in contrast to the conventional BCS case in which the even-frequency component of the pair amplitude contributes to the diamagnetism. 
We have further clarified that one of the $\eta$-pairing states on the triangular lattice has a stripe pattern and shows an anisotropic Meissner response.
In this case, the odd-frequency pair contributes diamagnetically or paramagnetically depending on the direction of current. 

We comment on some issues which are not explored in this paper.
We expect that the $\eta$-pairing without a simple staggered phase will appear on pyrochlore, kagome and quasicrystalline lattice, whose phase-alignment could be qualitatively different from the triangular lattice.
In addition, there is another model that shows $\eta$-pairing in equilibrium. A two-channel Kondo lattice (TCKL) is an example of a model in which $\eta$-pairing appears even in the absence of a Zeeman field \cite{Hoshino2014}. 
Our preliminary calculation for the TCKL shows a number of ordered states which have similar energies.
These additional studies provide more insight into the exotic superconductivity characteristic for the $\eta$-pairing.

\section*{Acknowledgement}

This work was supported by KAKENHI
Grants 
No. 18H01176, 
No. 19H01842, and 
No. 21K03459.

\begin{figure}[tb]
    \centering
    \includegraphics[width=8.6cm]{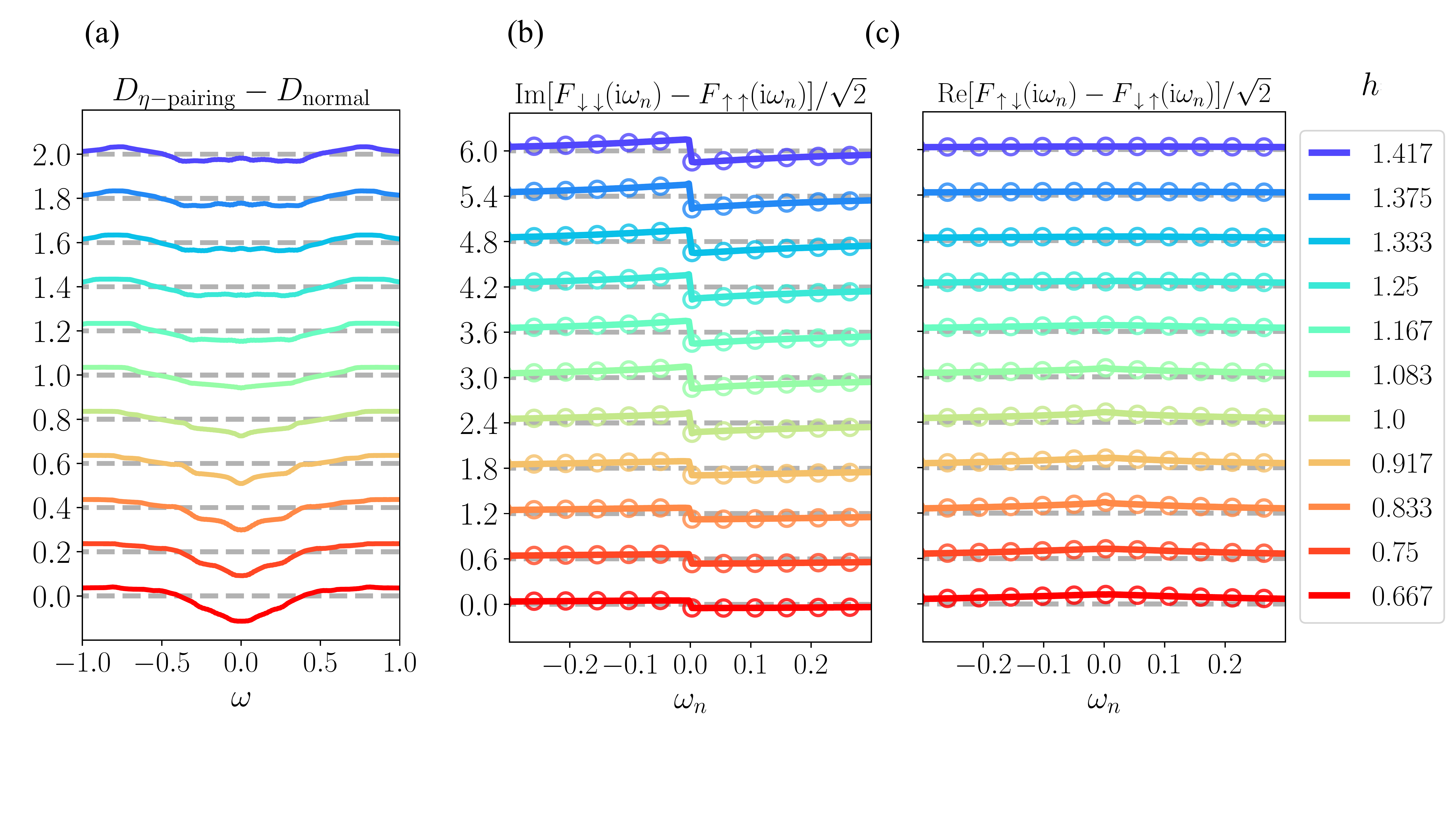}
    \caption{(a) The difference between the DOSs of the $\eta$-pairing and normal states in the cubic lattice model. The values of the DOS are shifted by 0.2 for each magnetic field, and the gray dotted lines are the zero axes for each magnetic field. We also show the  Matsubara frequency dependence of (b) the imaginary part of $\left[F_{\downarrow\downarrow}(\imu \omega_n) - F_{\uparrow\uparrow}(\imu \omega_n)\right]/\sqrt{2}$ and (c) the real part of $\left[F_{\uparrow\downarrow}(\imu \omega_n) - F_{\downarrow\uparrow}(\imu \omega_n)\right]/\sqrt{2}$ for each magnetic field. 
    The values of the pair amplitudes are shifted by 0.6.
    }
    \label{fig:DOS_pair_amplitude_cubic_lattice}
\end{figure}

\appendix
\section{Self-consistent equations in mean-field theory}
\label{sec:appendix_self}
We derive self-consistent equations for the general interacting Hamiltonian.
Let us begin with the Hamiltonian
\begin{align}
    \mathscr H &= \sum_{12} \ep_{12} c_1^\dg c_2
    + \sum_{1234} U_{1234} c_1^\dg c_2^\dg c_4 c_3
\end{align}
where site-spin indices are written as $1=(i_1,\sg_1)$.
The mean-field Hamiltonian is introduced as
\begin{align}
    \mathscr H_{\rm MF} &= \sum_{12} 
    \qty (E_{12} c_1^\dg c_2 
    + \Delta_{12} c_1^\dg c_2^\dg
    + \Delta_{12}^* c_2 c_1 ).
\end{align}
We assume $\la \mathscr H \ra = \la \mathscr H_{\rm MF} \ra$ where the statistical average is taken with $\mathscr H_{\rm MF}$.
Then the self-consistent equation is obtained as 
\begin{align}
    E_{12} &= \frac{\partial \la \mathscr H \ra}{\partial \la c_1^\dg c_2\ra}
    \nonumber \\
    &=\ep_{12} + 
    \sum_{34} 
    \qty(
    U_{1324} + U_{3142} - U_{1342} - U_{3124}
    )\la c_3^\dg c_4\ra
    \\
    \Delta_{12} &= 
    \frac{\partial \la \mathscr H \ra}{\partial \la c_1^\dg c_2^\dg\ra}
    = \sum_{34} U_{1234} \la c_4 c_3\ra
\end{align}
where the Wick's theorem is used for the derivation.
Although the variational principle for the free energy also gives the same equation, the above formalism gives a simple procedure to derive the self-consistent equations.

\section{Attractive Hubbard model on Cubic lattice}
\label{sec:appendix_cubic}
We analyze the $\eta$-pairing on the cubic lattice, whose DOS does not have a van Hove singularity near zero energy.
Here we choose the parameter $U=-1.375$ and the electron concentration is half-filled. As a result, the DOS for the $\eta$-pairing around zero energy for each magnetic filed on the cubic lattice is smaller than the DOS of the normal state as shown in Fig.~\ref{fig:DOS_pair_amplitude_cubic_lattice}(a). 
For reference, we also show in Figs.~ \ref{fig:DOS_pair_amplitude_cubic_lattice}(b) and (c) the pair amplitude similar to Fig.~\ref{fig:AH_Kernel_SQ}(b) in the main text. In addition, the odd-frequency pair amplitude increases when DOS near zero energy is enhanced as seen from Figs. \ref{fig:DOS_pair_amplitude_cubic_lattice}(a) and (b).

\end{document}